\documentclass[]{aastex}
\usepackage{emulateapj5,latexsym,graphics,epsfig}
\def \be{\begin{equation}}
\def \ee{\end{equation}}
\def \bea{\begin{eqnarray}}
\def \eea{\end{eqnarray}}
\def\etal{{et al.\ }}
\slugcomment{Submitted to ApJ ?}

\begin{document}

\title{Dust sputtering by Reverse Shocks in Supernova Remnants}

\author{Biman B. Nath$^1$, Tanmoy Laskar$^{1,2}$ and  J. Michael Shull$^3$}
\affil{$^1$Raman Research Institute, Sadashivanagar, Bangalore 560080, India\\
$^2$St. Stephen's College, Delhi 110007, India\\
$^3$CASA, Department of Astrophysical and Planetary Sciences, University
of Colorado, Boulder, CO 80309-0389, USA}
\email{biman@rri.res.in, mshull@casa.colorado.edu}

\shorttitle{Dust sputtering by reverse shock in SN}
\shortauthors{B. Nath, T. Laskar, J. M. Shull}

\begin{abstract}
We consider sputtering of dust grains, believed to be formed in cooling
supernovae ejecta, under the influence of reverse shocks. In the regime
of self-similar evolution of reverse shocks, we can follow the evolution
of ejecta density and temperature  analytically as a function
of time in different parts of the ejecta, and calculate the sputtering
rate of graphite and silicate grains embedded in the ejecta as they
encounter the reverse shock. Through analytic (1D) calculations, 
we find that a fraction of dust
mass ($ 1\hbox{--}20$\% for silicates and 
graphites)
can be sputtered by reverse shocks, the fraction varying with the
grain size distribution and the
steepness of the density profile of the ejecta mass. It is expected that
many more grains will get sputtered in the region between the forward and
reverse shocks, so that our analytical results provide a lower limit to
the destroyed fraction of dust mass.
\end{abstract}
 
\keywords
{ISM : Supernova Remnants, ISM : Dust, Extinction, Galaxies : High-Redshift}

\section{Introduction}
Understanding the chemical history of heavy elements (``metals")
in the interstellar medium (ISM) of galaxies is an important, but
difficult undertaking.  Because refractory elements are found to
be depleted from the gas phase (locked into grains), they provide
a substantial reservoir of coolants in the solid-state phase.
Interstellar grain surfaces are believed to furnish formation sites
for molecular hydrogen (H$_2$), which is both a coolant and the
starting point for a rich cloud chemistry. Grains also provide a
critical transfer mechanism for reprocessing UV/O starlight into
far-infrared and sub-mm emission from interstellar molecular clouds
and high-redshift galaxies.  In the high-redshift universe, the
first dust grains may influence the thermodynamics of the
primordial gas and thereby control rates of star formation.

Recent observations of dust grains at high
redshift have been puzzling in the context of their formation.
Several observations of damped Ly$\alpha$ systems (Pettini \etal 1994; Ledoux,
Bergeron \& Petitjean 2002) have shown evidence of dust grains in them. Also,
thermal emission (at $1.2$ mm) from dust in high redshift ($z >6$) QSOs in 
the Sloan Digital Sky Survey (SDSS)
have been reported (Bertoldi \etal 2003).
The implied dust mass of $\sim 10^8$ M$_\odot$
 within a Gyr after the Big Bang
 appears difficult to explain with traditional
models of grain formation in evolved low mass stars (Dunne \etal 2003), from
which grains are thought to be
transported to the interstellar medium (ISM)
through stellar winds (Whittet 1992).

Recent models have, therefore, focused on other sites
of dust grain formation, such as ejecta of core-collapse
 supernova explosions that
can occur on shorter time scales than that of the evolution of low mass stars
(Dwek \& Scalo 1980). Core-collapse supernovae are thought to be capable of
synthesizing a significant amount of 
dust in the SN ejecta.
Applying a theory of nucleation and grain growth developed by
Kozasa \& Hasegawa (1987), Kozasa \etal (1989; 1991)
 and Todini \& Ferrara (2001)
have studied  grain formation in expanding SN ejecta, the latter estimating
the dust mass formed $\sim 0.1\hbox{--}0.3$ M$_{\odot}$. Nozawa \etal (2003)
have extended it to the
case of zero-metallicity SNe for population III stars (see also Schneider,
Ferrara \& Salvaterra 2004).

Although the SN ``dust factory model'' would explain the large
observed dust masses in high-redshift galaxies, observations of
dust in nearby SNe appear inconsistent with this model, falling
factors of 10--100 short of the required amounts ($\sim0.2M_{\odot}$;
see above).
As an example,
in a review of Type~II SN~2003gd (NGC~628),
Sugerman et al.\ (2006) state that ``radiative transfer models show
that up to 0.02 solar masses of dust has formed within the ejecta".
However, a recent study (Meikle et al.\ 2007) of the same remnant
concludes that the mid-IR flux ``is consistent with emission from
$4 \times 10^{-5}~M_{\odot}$ of newly condensed dust in the ejecta".
In a young, oxygen-rich SNR in the SMC, Stanimirovic et al.\ (2005)
found only $10^{-3}~M_{\odot}$ of hot dust ($T_d \approx 120$~K).
The claimed sub-mm detection (Dunne et al.\ 2003) of $\sim3~M_{\odot}$
of cold dust toward the Cas~A remnant was later shown to arise from
interstellar dust in an adjacent molecular cloud (Krause et al.\ 2004).

Observational tests of predictions made by models of 
SNe are difficult, involving far-IR and
sub-mm observations of young SNRs.  One must disentangle far-IR
backgrounds, separate newly synthesized dust from circumstellar
and interstellar dust, and understand the role of SNR reverse
shocks, which can destroy newly synthesized dust in high-speed ejecta.
A case in point is SN 1987A.
Dust formation was inferred in ejecta of SN~1987A, but only
$10^{-3}~M_{\odot}$ was detected (Dwek 2006; Dwek \& Arendt 2007).
Additional dust is almost certainly present, because of observed changes
in the optical and bolometric fluxes and emission-line asymmetries in
the red and blue wings (McCray 1993, 2007).  Moreover, the reverse shock
has not yet reached the interior of the supernova debris in SN~1987A.
These inner ejecta are most prone to dust formation, since they are
cold, dense, and metal-enriched.

A challenging theoretical issue is whether high-velocity dust
($V_{\rm ej} \geq 1000$ km~s$^{-1}$) can survive the shocks,
thermal sputtering, and grain-grain collisions, as the ejecta
are slowed down by the surrounding ISM.
A goal of this paper is to determine how much dust survives
the SN event and is incorporated into the surrounding ISM. The relative
speed of reverse shocks with respect to the expanding ejecta can be very large,
and can raise the temperature of the ejecta to the extent of destroying the
nascent dust grains in it. It is therefore important to study the effect
of reverse shocks in detail.

Recently, Nozawa \etal (2007) have considered this problem and came to
the conclusion, with the help of hydrodynamical simulation, that reverse
shocks can destroy a fraction of dust mass of order $20\hbox{--}100$\%
depending on the ambient gas density and explosion energy. Bianchi \&
Schneider (2007) also used numerical simulations and semi-analytical methods to
conclude that only a small amount of dust mass survives; e.g.,
 $\sim 7$\% dust mass survives in the case of a SN with
a progenitor of mass $20 \,$ M$_{\odot}$ and ambient particle density of
$\sim 1$ cm$^{-3}$. Nozawa \etal (2007) have included the motion
of dust relative to gas within the SN remnant and calculated the evolution
of dust grains over a period spanning the radiative phase ($\sim 10^5 \hbox{--}
10^6$ yr), whereas Bianchi \& Schneider (2007) neglected the effect
of motion of grains on dust destruction and calculated dust evolution
up to the non-radiative phase ($\le 10^5$ yr).

We study in this paper a simple and analytically tractable problem, namely
 the effect of reverse shocks for a power-law
 mass distribution of the ejecta, in the
regime of self-similar evolution of both forward and reverse shocks. Although
the simple formalism does not allow us to study several important
processes--- including motion of dust grains and its effect on evolution
of dust grains--- our approach allows one to segregate the processes
of destruction within the reverse shock itself from those occuring between
forward and reverse shocks (which can be dealt with in numerical
simulations involving grain motion).

\section{Dust in supernova remnants}
We first summarize the results of the theoretical predictions of dust grain
formation in SNRs. Todini \& Ferrara (2001) calculated the formation of
different types of dust grains--- amorphous carbon (AC), silicates (enstatites
: $MgSiO_3$, forsterites : $Mg_2SiO_4$) and iron-bearing magnetites ($Fe_3O_4$)
and others, in ejecta with different metallicities. They found that AC
grains are generally of larger size, in the range of $\sim 0.01\hbox{--}0.1 \,
\mu$~m, whereas silicates are produced with sizes $\sim 10^{-3} \, \mu$~m.
Nozawa \etal (2003) reached similar conclusions for Pop III SN events, with
sizes of newly synthesized grains spanning a range of three orders of
magnitude, with maximum grain size being less than  $\sim 1 \, \mu$m.

The dust grain parameters in the ISM are inferred to be somewhat
different from these
estimates. The extinction observations are satisfactorily reproduced by
assuming mostly of two components--- graphite and silicate--- with
a common power-law size distribution, $dn/da \propto a^{-3.5}$, truncated
at a minimum size $a_l\sim 5 \times 10^{-3} \, \mu$m  and a maximum size
$a_m \sim 0.25 \, \mu$m (Draine 2003, and references therein). There
also appears to be a population of very small grains with PAH composition.
Nozawa \etal (2003) found for grains synthesized in Pop III SN ejecta
a size distribution with index $-2.5$ for smaller grains and with $-3.5$ for
larger grains (with the grain size at crossover point depending on supernovae
models). We use these two power laws to bracket the possible size
distributions of grains synthesized in SN ejecta, and assume that grains
are formed in the range $10^{-7}\hbox{--}3 \times 10^{-5}$ cm (see Figure 11 of
Nozawa \etal (2003). We also consider the case of an upper limit of
$10^{-6}$ cm and compare our results for both upper limits.

It is instructive to estimate the destruction time scales before we calculate
the effect of reverse shock in detail.
 The sputtering time scale for dust (graphite,
silicate and iron) grains of size $a$
 is approximately $\sim 10^6 \, ({a \over 1 \, \mu {\rm
m}} ) \, ({n_e 
\over 1 \, 
{\rm cm}^{-3}})^{-1}$ yr, for ambient temperatures larger than $10^6$ K
(Draine \& Salpeter 1979), where  $n_e$
is electron density. For grains with size $\sim 0.1\hbox{--}0.3$ $\mu m$
(which incidentally contain most of the dust mass for a $a^{-3.5}$
distribution), and for $n_e\sim 1\hbox{--}10$ $cm
^{-3}$, the sputtering time scale is $\sim 2 \times 10^{4\hbox{--}5}$ yr.

\section{Reverse shocks}
We now briefly  discuss the self-similar evolution of
forward and reverse shocks in SN (see Truelove \& McKee 1999, hereafter
referred to as TM99, for further
details).

\subsection{The ejecta}
The parameters used to describe a supernova remnant which affect its
evolution are
the energy of the explosion ($E_{ej}$),
the mass contained in the expanding ejecta ($M_{ej}$),
the maximum velocity of the material within the ejecta ($v_{ej}$), and
the density of the ISM ($\rho_{0}$).
Denoting the time of the explosion as $t=0$, we construct our
initial conditions at a later time, still early in the history of the remnant,
assuming  the ejecta to have already expanded to a radius $R_{ej}$
without any deceleration due to the ambient medium.
If we consider SN to be a point explosion, then the radial velocity profile
will be linear as ejecta with velocity $v$ will travel out to a distance $vt$ :
\begin{equation}
 v(r) = \left\{ \begin{array} {ll}
                  \frac{r}{t}, & \mbox{$r < R_{ej}$} \\
                  0, & \mbox{$r > R_{ej}$}
                  \end{array}
            \right .
\end{equation}
This profile becomes flatter as time progresses and the faster-moving
ejecta are flung out to greater distances.
The corresponding radial density distribution of the ejecta matter is given by:
\begin{equation}
\rho(r) = \left\{ \begin{array} {ll}
                  \rho_{ej}(v,t) \equiv \frac{M_{ej}}{v_{ej}^3} 
f(\frac{v}{v_{ej}})t^{-3}, & \mbox{$r < R_{ej}$} \\
                  \rho_{0}, & \mbox{$r > R_{ej}$}
                  \end{array}
            \right.
\label{eq:den}
\end{equation}
where $f(v/v_{ej})$ is the time-independent structure function of the density
profile and the $t^{-3}$ term arises from the free expansion of the ejecta.
We define a dimensionless parameter to label shells of the ejecta starting
from the centre outwards as
\be
w = \frac{v}{v_{ej}},        0 \leq w \leq 1
\ee
Following TM99 we consider the situation when
\textit{f} is given by a power law in \textit{w}:
\be
f(w) = \left\{ \begin{array} {ll}
                  f_{0}, & \mbox{$0 \leq w \leq w_{c}$} \\
                  f_{n}w^{-n}, & \mbox{$w_{c} \leq w \leq 1$}
                  \end{array}
            \right.
\ee
where we have separated the ejecta into a uniform density core region
and a power-law envelope region.  The continuity of \textit{f} at $w_{c}$
and the normalization of $\rho$ translate to expressions for parameters $f_{0}$
and $f_{n}$ in terms of the free parameters $w_{c}$ and \textit{n}:
\bea
f_{0} &=& f_{n}w_{c}^{-n} \, \nonumber\\
f_{n} &=& \frac{3}{4\pi} \left[ \frac{1-n/3}{1-(n/3)w_{c}^{3-n}}\right] \,.
\eea

\subsection{Forward and reverse shocks}
As the freely expanding ejecta come into contact with the ambient medium, with
the contact discontinuity moving at $v_{ej}$, a speed much larger than the
sound speed in the ambient medium, a forward shock is set up in the
surrounding gas
which propagates into the medium. The ejecta at the contact surface decelerates
suddenly as it hits the ambient gas, and consequently, a reverse shock is
set up that propagates inwards through the ejecta. While the mass of this
matter swept by the forward shock is less than the total mass of the ejecta
driving the shock, the remnant is said
to be in the \textit{Ejecta Dominated} or ED stage. It is in this stage, before
 the reverse shock has attained a significant velocity, that the ejecta can be
approximated as a spherical piston freely expanding into the ambient gas. When
the mass swept by the forward shock becomes approximately equal to the ejecta
mass, the corresponding late-time limit is known as the
\textit{Sedov-Taylor} or ST stage of the evolution of the remnant.

We follow TM99 to follow the evolution of the reverse shock analytically.
Let $R_{f}$ and $R_{r}$ denote the radii of the forward and reverse
shocks, and $v_{f} \equiv dR_{f}/dt$ and
$v_{r} \equiv dR_{r}/dt$ be the velocities of the forward and reverse shock
waves respectively, all in the rest frame of the unshocked ambient medium.
We have the velocity of the unshocked ejecta as seen by the reverse shock as
it propagates into the ejecta:
\be
\tilde{v}_{r} \equiv v(R_{r},t) - v_{r}(t) = \frac{R_{r}(t)}{t} - v_{r}(t)\,.
\ee
Let \textit{$\phi$} be the ratio of the pressures behind the reverse and
forward shock waves and \textit{l} be the ratio of the radii of the two shocks
at a given time:
\be
\phi(t) \equiv \frac{\rho_{ej}(v_{r},t)\tilde{v}_{r}^2(t)}{\rho_{0}v_{f}^2(t)}
\,,\quad
l(t) \equiv \frac{R_{f}(t)}{R_{r}(t)} \,, \quad
w_f(t) \equiv \frac{R_f(t)/t}{v_{ej}}  \,.
\ee

It is useful to express various parameters in dimensionless forms in the
units of characteristic values of these variables. so that,
$R_\ast=R/R_{ch}$,
$t_\ast=t/t_{ch}$, and  $v_\ast=v/v_{ch}$. For example, the
characteristic scales of length, time and velocity are given by,
\bea
R_{ch}&=& M_{ej}^{1/3} \rho_0^{-1/3} \approx 5.25 \,   n_0^{-1/3} \,
\, {\rm pc} \,, \nonumber\\
t_{ch}&=& E^{-1/2} M_{ej}^{5/6} \rho_0^{-1/3}
\approx 1.66 \times 10^3   \, n_0^{-1/3} \,\, {\rm yr} \,, \nonumber\\
v_{ch}&=&  R_{ch}/t_{ch} \approx 3.16 \times 10^3 \, {\rm km} \,\,
{\rm s}^{-1}\,.
\label{eq:char}
\eea
(We have assumed $\rho _0=1.4 n_0 m_p$ for He/H$=0.1$ by number.)
We have estimated the numerical values for the case $E=10^{51}$ erg, $M_{ej}=
10^{34}$ g and $n_0=1$ cm$^{-3}$.
We will therefore write
$R^\ast=R/R_{ch}$,
$t^\ast=t/t_{ch}$, and  $v^\ast=v/v_{ch}$, and the
radius of the reverse shock, in particular, as
$R_r^\ast\equiv R_r/R_{ch}$.

\subsubsection{ED (Expansion dominated) stage}
The values of $\phi$ and $l$ approach constant values as $t \rightarrow 0$ if
the solutions remain self-similar in the same limit. We write $\phi(t) \simeq 
\phi(0) \equiv \phi$ and $l(t) \simeq l(0) \equiv l$, and
also $w_{f}(0) = l$. The solution to reverse shock evolution exists in the form
 of two branches, which we label the \textit{envelope} ($w_{c} \leq 
w_{f}/l_{ED} \leq 1$) and \textit{core} ($0 \leq w_{f}/l_{ED} \leq w_{c}$)
branches. Denoting $\alpha \equiv \frac{E}{(1/2)M_{ej}v_{ej}^{2}}$,
 the evolution of the reverse shock is given
by,
\bea
&&t^\ast(R_{r}^\ast) = \left(\frac{\alpha}{2}\right)^{1/2}R_{r}^\ast \times
\nonumber\\
&&
 \left\{ \begin{array} {ll}
\left[1 + \left(\frac{n-3}{3}\right)\left(\frac{\phi}{f_{n}}\right)^{1/2} l
R_{r}^{\ast 3/2}\right]^{-2/(3-n)}
, \\
\left[w_{c} - \left(\frac{\phi}{lf_{0}}\right)^{1/2} \left\{l^{3/2}R
_{r}^{\ast 3/2} - \frac{f_{n}^{1/2}}{3-n}\left(1 - w_{c}^{(3-n)/2}\right)
\right\}\right]^{-2/3}
, 
                  \end{array}
            \right.
\label{eq:ednl3_1}
\eea
where the first part corresponds to the case $w_{c} \leq w_{f}/l \leq 1$
and the second part, to $0 \leq w_{f}/l \leq w_{c}$. The corresponding
 velocity of reverse shock in these two cases are,
\bea
&& v^\ast(R_{r}^\ast) = \left(\frac{2}{\alpha}\right)^{1/2} \times \nonumber\\
&& \left\{ \begin{array} {ll}
\frac{
\left[1 - \left(\frac{3-n}{3}\right)\left(\frac{\phi}{f_{n}}\right)^{1/2}
lR_{r}^{\ast 3/2}\right]^{(5-n)/(3-n)}}{\left[1+\frac{n}{3}\left(\frac{\phi}
{f_{n}}\right)^{1/2}lR_{r}^{\ast 3/2}\right]}
, & \\

\frac{\left[w_{c} - \left(\frac{\phi}{lf_{0}}\right)^{1/2} \left\{l^{3/2}
R_{r}^{\ast 3/2} - \frac{f_{n}^{1/2}}{3-n}\left(1 - w_{c}^{(3-n)/2}\right)
\right\}\right]^{5/3}}{\left[w_{c} + \frac{f_{n}^{1/2}}{3-n}\left(1 -
w_{c}^{(3-n)/2}\right)\right]}
, & 
                  \end{array}
            \right.
\eea
The relative velocity of the reverse shock in the frame of reference of
the expanding ejecta, for the case when the reverse shock is in the
envelope region, is given by,
\be
\tilde{v}_r^\ast(R_{r}^\ast) = %
\sqrt{{2 \phi \over \alpha f_{n}}}
lR_{r}^{\ast 3/2} \,
\left(
\frac{
\left[1 - \left(\frac{3-n}{3}\right)\left(\frac{\phi}{f_{n}}\right)^{1/2}
lR_{r}^{\ast 3/2}\right]^{2/(3-n)}}{\left[1+\frac{n}{3}\left(\frac{\phi}{f_{n}}
\right)^{1/2}lR_{r}^{\ast 3/2}\right]}\right) \,.
\label{eq:ednl3_2}
\ee
Note that, since $w_f/l =\frac{R_r(t)/t}{v_{ej}}$
describes the reverse shock, we will define it as `$w$' and use it to label
shells in the ejecta.

This solution is valid (in the ED stage) only as long as
the reverse shock remains in the envelope $w_{core}\leq w \leq 1$.
In the case of $n < 3$ ejecta, there is no need for a core and we can take the
limit $w_{core} \rightarrow 0$. For $n > 3$ however, we must have a ``flat
profile'' core to keep the integrated mass finite.
For the steep index scenario, therefore, the functional form of the solution
will change when the reverse shock reaches the core and we will return to
this point later.

\subsubsection{ST (Sedov-Taylor) stage}
We now describe the Sedov-Taylor stage, which is the late-time evolution limit
of the non-radiative phase of a supernova remnant.
Following TM99, we assume that the shocks trace a constant acceleration path
after transition to the ST stage at $t_{ST}$, which (approximately) begins
when the mass of the ambient medium shocked by the forward shock equals the
mass ejected by the supernova. Thus the velocity of the oncoming ejecta in the
frame of the reverse shock is
\be
\tilde{v}_{r}^{*}(t^{*}) = \tilde{v}_{r,ST}^{*} + \tilde{a}_{r,ST}^{*}(t^{*}
- t_{ST}^{*}) \,,
\label{eq:stnl3_2}
\ee
where $\tilde{a}_{r,ST}$ is the constant acceleration and $\tilde{v}_{r,ST}$
is the velocity of the ejecta w.r.t the reverse shock at $t_{ST}$.
Note that $\tilde{v}_{r} = -td(R_{r}/t)/dt$. Using this and integrating the resulting expression from $t_{ST}$ to $t$ gives
\be
R_r ^\ast (t^\ast) = t^\ast \left[\frac{R_{r,ST}^\ast} {t_{ST}^\ast} -
\tilde{a}_{r,ST}^\ast (t^\ast -t_{ST}^\ast ) - (\tilde{v}_{r,ST}^\ast  -
\tilde{a}_{r,ST}^\ast t_{ST}^\ast )
 ln\left(\frac{t^\ast }{t_{ST}^\ast }\right)\right] \,.
\label{eq:stnl3_1}
\ee

For the $n < 3$ case, we therefore have an implicit solution for the reverse
shock position in the ED stage $t < t_{ST}$ and an explicit solution in the
form of a uniform acceleration description in the ST stage.
For the $n > 5$ case, we again have an implicit solution in the ED stage at
least while $t < t_{core}$. In fact, an explicit solution for $R_{r}^\ast (t^
\ast )$ can be determined from the implicit solution in the ED stage for
$n > 5$ remnants. We have (see equation (74) of TM99),
\be
t^\ast (R_{r}^\ast ) = \left(\frac{\alpha}{2}\right)^{1/2}R_{r}^\ast \,
\left[1 + \left(\frac{n-3}{3}\right)\left(\frac{\phi}{f_{n}}\right)^{1/2}
lR_{r}^{\ast 3/2}\right]^{-2/(3-n)} \,.
\ee
In this expression, we can take the limit $w_{core} \rightarrow 0$ by
allowing $v_{ej} \rightarrow \infty$, the ejecta energy $E$ remaining finite.
In this limit $f_{n} \rightarrow 0$ and when the second term in the brackets
becomes much larger than unity, we have
\be
R_{r}^\ast (t^\ast ) = \left[\frac{27}{4\pi}\frac{1}{n(n-3)l^{2}\phi}
\left\{\frac{10}{3}\left(\frac{n-5}{n-3}\right)\right\}^{\frac{n-3}{2}}\right]
^{1/n} \bigl ( t^{\ast} \bigr ) ^{ \frac{n-3}{n}} \,.
\label{eq:tlc_n5_1}
\ee
This gives us the velocity of the reverse shock as,
\be
{v}_{r} ^\ast (t^\ast ) =
\frac{n-3}{n} \, \left[\frac{27}{4\pi}\frac{1}{n(n-3)
l^{2}\phi}\left\{\frac{10}{3}\left(\frac{n-5}{n-3}\right)\right\}^{\frac{n-3}
{2}}\right]
^{1/n} \, \bigl (t^{\ast} \bigr ) ^{ -3/n} \,,
\ee
and correspondingly,
\bea
\tilde{v}_{r}^\ast (t^\ast )&=&\frac{R_{r}}{t} - v_{r}^\ast \nonumber\\
&=&
\frac{3}{n}\left[\frac{27}{4\pi}\frac{1}{n(n-3)l^{2}\phi}\left\{\frac{10}
{3}\left(\frac{n-5}{n-3}\right)\right\}^{\frac{n-3}{2}}\right]^{1/n}
t^{\ast -3/n} \,.
\label{eq:tlc_n5_2}
\eea
This is the result arrived at by Chevalier (1982) and Nadyozhin (1985) -
henceforth referred to as `the CN solution'. The transition of a supernova
remnant with $n > 5$ to this stage has been shown to occur extremely fast
(TM99).
Thus for all practical purposes,
we may assume that the remnant begins with the CN stage.

The reverse shock then moves into the core at $t = t_{core}$.
 TM99 assumed
that the reverse shock motion after this time is also described by a
Sedov-Taylor trajectory - which leads to better match between the two regimes
in terms of continuity of the shock radius and velocity.
 In the case of $n > 5$, therefore, if we replace $t_{ST}$ by $t_{core}$ in
the ST stage equations above, the same analysis follows through.
 The numerical values of the parameters $t_{core}, \tilde{a}_{r,core}$ and
$\tilde{v}_{r,core}$ are, of course, different from the parameters for the
$n < 3$ case: viz $t_{ST}$ and $\tilde{a}_{r,ST}$.
We have,
\bea
R_r^\ast (t^\ast ) &=& t^\ast \Bigl [ \frac{R_{r,core}^\ast }{t_{core}^\ast }
- \tilde{a}_{r,core}^\ast (t^\ast -t_{core}^\ast ) \nonumber\\
&& - (\tilde{v}_{r,core}^\ast
 - \tilde{a}_{r,core}^\ast t_{core}^\ast ) ln\left(\frac{t^\ast }{t_{core}^\ast
}\right) \Bigr ] \nonumber\\
\tilde{v}_r^\ast (t^\ast ) &=& \tilde{v}_{r,core}^\ast  + \tilde{a}_{r,core}
^\ast (t^\ast  - t_{core}^\ast ) \,.
\label{eq:tmc_n5_1}
\eea
The values of the parameters used in our calculations have been taken from TM99,
 which have been checked with results from simulations.

\subsubsection{Summary of ED and ST solutions}
We thus have the final solutions for the reverse shock
as follows. In the case of $n < 3$,
equations (\ref{eq:ednl3_1}; first part) and (\ref{eq:ednl3_2}) describe the
solutions in the ED stage, and, thereafter, equations (\ref{eq:stnl3_1}) and
(\ref{eq:stnl3_2}) are appropriate for solutions in the ST stage.
For the case of $n>5$, solutions are given by
equations (\ref{eq:tlc_n5_1}) and (\ref{eq:tlc_n5_2})
in the $t < t_{core}$ stage. Afterwards,
in the $t > t_{core}$ stage, equations (\ref{eq:tmc_n5_1})
describe the solutions. The analytical solutions remain valid
until $t_\ast \sim 2.2$ (TM99), and we calculate the evolution
of shocks until this point (except for $n=4$; see below).

For a given set of values of $E$ and $M_{ej}$, we can therefore
solve for reverse shock position and relative velocity, assuming a
density distribution described by index $n$. The parameter $\alpha$
in different cases of $n$, is given by,
\be
\alpha\equiv {E \over (1/2) M_{ej} v_{ej}^2}=\Bigl ( {3 -n \over 5-n} \Bigr )
\Bigl ( {w_{core}^{-(5-n)}-n/5 \over w_{core}^{-(3-n)}-n/3} \Bigr )
w_{core}^2 \,,
\ee
For $n<3$, we calculate this in the limit of $w_{core}=0$, and for
$n>5$ (including the case of $n=4$)
 we assume (following TM99) a value of $w_{core}=0.1$. Values
of other parameters (such as $l, \phi$) are taken from appropriate tables
of TM99 for the relevant values of $n$.

We also consider the special case of $n=4$, for which the mass integral
diverges without a core. We follow the prescriptions of TM99 (their section 8)
for this case (and, again, assume $w_{core}=0.1$),
and track the reverse shock with only ED solution without
any transition to the ST case. According to TM99, based on their comparison
of analytical and numerical results, this prescription gives tolerably good
results (accurate to within $24\%$) until $t^\ast =1.2$.

\subsection{Arrival time of reverse shock}
We have introduced a label to a given shell in the supernova ejecta by $w
\equiv w_f/l =\frac{R_r(t)/t}{v_{ej}}$, the ratio of the velocity of the ejecta
in the shell to the maximum ejecta velocity. The ejecta in a shell specified by
 a given $w$ first undergoes free expansion (in the ED stage) and
 self-similar expansion (in the ST stage) until hit by
the incoming reverse shock wave. This occurs at a time, which we call
$t_{hit}$, which is a function of $w$ and which is found by equating $w$ to
the $v_{r}/v_{ej}$
found from the position of the reverse shock. That is,
\be w = \frac{v_{r}}{v_{ej}} = \frac{R_{r}}{R_{ej}}
\Rightarrow R_{r} = w R_{ej} = w v_{ej} t_{hit} \,.
\ee
Or, in terms of dimensionless variables,
\be
R_{r}^\ast  = w\cdot \frac{v_{ej}t_{hit}}{R_{ch}} =  \left( w\cdot \frac{v_{ej}
t_{ch}}{R_{ch}} \right) t_{hit} ^\ast \,.
\ee
This expression for $R_{r}^\ast $ in terms of $t^\ast $ is equated to the
explicit solution  for the reverse shock, when available, as a function of time,
and the resultant linear equation in $t_{hit}^{*}$ and $w$ is solved using
the bisection method for each given $w$. This is applicable to the steep
($n > 5$) ejecta case (both the CN and ST stage) as well as to the ST stage
for shallow ($n < 3$) ejecta. For the ED stage of shallow ejecta, we use
 $t_{hit}^{*}$\
 as a function of $R_{r}^{*}$ in equation (\ref{eq:ednl3_1}) and then
employ the bisection method exactly as before to solve for $t^\ast _{hit}$.
Either way, we can determine $t^\ast _{hit}$ given any shell labeled
by a $w$ in the SN ejecta.

\subsection{Density and temperature evolution in ejecta shells}
We assume that gas in a given shell of the ejecta evolves adiabatically before
$t_{hit}$. The temperature is then raised to the postshock temperature and the
density acquires a jump, corresponding
to the value of relative shock velocity $\tilde{v}_r$. After $t_{hit}$, the
gas again evolves adiabatically, that is, its density decreases as $t^{-3}$
(equation \ref{eq:den}). Its temperature decreases as $T \propto \rho^
{\gamma -1} \propto t^{-2}$, for $\gamma =5/3$ appropriate for monoatomic gas.
For a shell with a certain value of $w$, therefore, we have the following
evolution of density,
\be
\rho(t,w) = \left\{ \begin{array} {ll}
                  \frac{M_{ej}}{v_{ej}^3}
f(w)t^{-3}, & \mbox{$t < t_{hit}$} \\
            4  \frac{M_{ej}}{v_{ej}^3} f(w) t^{-3}
                  , & \mbox{$t_{hit} \le t < 2.2 t_{ch}$}
                  \end{array}
            \right.
\label{eq:den_ev}
\ee
and temperature,
\be
T(t,w) = \left\{ \begin{array} {ll} T_i (\frac{t}{t_i})^{-2},
& \mbox{$t < t_{hit}$} \\
T_{rshock} (\frac{t}{t_{hit}})^{-2},
                  , & \mbox{$ t_{hit} \le t < 2.2 t_{ch}$}
                  \end{array}
            \right.
\ee
where $T_{rshock}=\frac{3}{16} \mu m_p \tilde{v_r}^2/k_B$, the post-shock
temperature due to the reverse shock with relative velocity $\tilde{v_r}$.
For a completely ionized gas the mean molecular weight $\mu=0.6$.
 We assume a value of $T_i=5400$ K at $t_i=70$ days, motivated by 
photometric observations of SN1987A (Catchpole \etal 1987) where it was
estimated that the photospheric temperature at $\sim 70$ d after the
explosion was $\sim 5400$ K. We have also checked that our results do not
depend strongly on these assumptions.

\subsection{Sputtering of dust}
Given the above mentioned evolution of density and temperature in shells
of ejecta labeled by different values of $w$, we can compute the steady
decrease in grain radius, for different types of dust grains. We use the
polynomial fit given by Tielens et al (1994) for  graphites and silicates
as a function of gas temperature. They expressed the rate, $(1/n_H) {da \over
dt}$ in powers of $\log _{10} T$, with coefficients for different grain
composition tabulated in their
Table 4 (see their equation 4.21).
These rates are consistent with recent results of Nozawa \etal  (2006).

We follow a shell of ejecta material labeled by a
value $0 < w < 1$ from
$\sim$ 10 years after the explosion (by which time dust grains are
believed to have formed in the cooling ejecta, since the formation time
scale is of order a few years (see, e.g., Todini \& Ferrara 2001)),
to when the reverse shock hits the shell at $t_{hit}$ and  beyond, until a
time $t^\ast  \sim 2.2$, which is the limit of the validity of the analytical
solutions.

\section{Results}
We have calculated the effect of reverse shocks on sputtering of graphite
and silicate grains for typical value of explosion energy, stellar ejecta
mass and ambient density. Below we show the results for $E=10^{51}$ erg,
$M_{ej}=10^{34}$ g, $\rho_0=10^{-24}$ g cm$^{-3}$.

First, we show  the evolution of grain radius (for graphites and silicates) for
two dust grains embedded in two particular shells, for example, in ejecta
characterized by $n=0$ and $E=10^{51}$ erg. We show in Figure 1 the cases
for the shells marked $w=0.3$, and $0.6$. The reverse shock passes
through these two shells, first hitting the $w=0.6$ shell at $\sim 0.5
t_{ch} \sim 0.8 \times 10^3 n_0^{-1/3}$ yr, and then
the $w=0.3$ shell at $\sim 1.2
t_{ch} \sim 2 \times 10^3 n_0^{-1/3}$ yr. The bottom panel of Figure 1
shows 
the rise in temperature for these two shells followed by adiabatic cooling.
Initially the temperature drops to very low temperatures due to strong
adiabatic cooling, although,
in reality, gas cannot be cooled below the temperature of the cosmic
microwave background, but we have not used such a lower bound in our
calculations. At any rate, the post-shock temperature does not depend on the
low temperature of the gas upstream, being determined only by the strength
of the reverse shock, and so the results of our calculation
remains realistic in spite of the strong cooling before the reverse shock hits
the gas in a given shell.

Figure 2 plots the ratio of final to initial grain sizes as a function
of the initial grain sizes for silicates and graphites forming (and being
sputtered) in shells $w=0.2, 0.4, 0.6$ for the particular case of $n=0$.
{\it Since the sputtering yield of silicate grains is larger than that of
graphites for a given set of parameters}, silicates are eroded more rapidly
 than graphite grains for sputtering
in a given ejecta shell. Also, the relatively outer shell of $w=0.6$
experiences less sputtering than the inner shells of $w=0.2$ and $0.4$. This
is expected since the reverse shock (relative) speed picks up as it plows
through the ejecta, increasing with decreasing values of $w$.
To illustrate this point,
we plot the trend of the relative velocity of the reverse shock as a function
of $w$ for the shells hit by the shock in Figure 3, for $n=0$ (solid),
$n=2$ (dotted), and $n=6$ (dashed). 
The relative
velocity is a rising function with decreasing  $w$, for $n=0$ and
for $n>5$, and it is a peaked function for $n=2$ (and also for $n=4$, which
is not shown here, but as shown by TM99).

However, we notice in Figure 2 that grains (both graphites and silicates)
in shell $w=0.2$ (solid curve) are sputtered to a somewhat
lesser extent than grains in the $w=0.4$ shell, although the relative speed
of the reverse shock  should be (according to Figure 3, solid curve for $n=0$)
higher when it hits the $w=0.2$ shell than at the $w=0.4$ shell.
Although the post-shock temperature at $w=0.2$ is higher than at $w=0.4$,
 the shells in the inner region
encounter the reverse shock later
 than shells in the outer region of ejecta,
and by the time they encounter it, the density in the inner shells
would have decreased rapidly (see equation \ref{eq:den_ev}). The rarefaction
in the inner shells at the time of encounter with the reverse shocks
lessens the effect of dust sputtering in these shells.

We illustrate this point by considering grains which are sputtered below
a certain size, say, $10^{-7}$ cm ($10^{-3} \, \mu$m),
as a function of the shell in which
they reside (and get sputtered).
We determine the
maximum size of grains which are sputtered below a size $\sim 10^{-7}$ cm,
 as a function of $w$, for different shells, and present the results in
Figure 4.
Curves for different
values of $n$ show the maximum size of grains sputtered below $0.001 \, \mu$m
 as a function of the parameter $w$ (left panels), as well as against
the fraction of ejecta mass that is
contained inside of the concerned shell, $M(<w)/M_{ej}$ in the right panels.
Results for
graphites and silicate grains  are shown in the upper and lower
panels respectively.

Figure 4 clearly shows that sputtering becomes intense in intermediate
shells, and decreases (as a result of rarefaction) with decreasing value
of $w$ for small values of $w$.
The sizes of grains
which are sputtered below $10 ^{-3} \mu $m
lie in the range of $\le 10^{-3} \, \mu$m 
$\hbox{--}10^{-2.6} \, \mu$m, with silicates more
strongly sputtered than graphites. In all cases,
the curves of
the maximum
size of sputtered grains
 increases with decreasing  values of $w$ and $M(<w)/M_{ej}$, finally
plunging to very low values for $w \ll 1$,
indicating negligible effect of the reverse shock deep
inside the ejecta.
For example, the (dotted)
curve for $n=2$ shows a peak at $w \sim 0.17$ below which the maximum
size of sputtered grains decreases.

Also, although it is not explicitly shown
in Figure 4, the corresponding results for $n >5$ show that reverse shocks
do not penetrate most of the ejecta mass in the time period of the
validity range
of analytical solutions $t_\ast \sim 2.2$.
Analytical means allow us to
probe deep into the ejecta mass distribution only for $n < 3$, and
full hydrodynamical models are needed to address the cases of steeper
profiles. For steep
density distributions,  the limiting
time of $t=2.2t_{ch}$ is reached by the time the reverse shock only skims
the surface of the ejecta, as far as mass fraction is concerned. The
characteristic time in the case considered here ($E=10^{51}$ erg, $M_{ej}=
10^{34}$ g, ambient density $\rho_0=10^{-24}$ g cm$^{-3}$)
is $t_{ch} \sim 1.7 \times 10^3$ yr. Although, by the limiting
time of $2.2 t_{ch} \sim 3.7 \times 10^3$ yr for the validity of self-similar
regime, reverse shocks can hit shells with small $w$ in these steep cases,
the mass contained inside of these shells remains very large.
 In other words, in
these cases we cannot probe what happens to most of the ejecta mass, or the
dust contained in them with analytical means, within the limits of
the self-similar regime of shock evolution.

We also note that a substantial amount of grain sputtering  takes place
in the hot gas between the forward and reverse shock that analytical
methods cannot capture. Detailed hydrodynamical calculations
are needed to address these issues, and our result can only provide a
lower limit to the total mass fraction of dust that is destroyed.

Finally, we calculate the total dust mass sputtered away as a function of
initial grain sizes in the following manner. We first calculate for a given
shell $w$, the final grain size $a_f (a_i,w)$ corresponding to the
 initial grain
size $a_i$. Then the fraction $1-(a_f/a_i)^3$ is the total dust mass sputtered
for grains formed in this shell. We then calculate the total dust mass
sputtered for all shells (for a given initial grain size $a_i$),
weighted by the shell mass in which they are situated, as given by,
\be
f_d (a_i)= \int  \bigl [ 1-({a_f \over a_i})^3 \bigr ] \, {d M(<w) \over
M_{ej}} \,,
\ee
where $M(<w)$ is the ejecta mass contained within the shell $w$.

We plot these weighted fractions $f_d (a_i)$ for graphite and silicate grains
as functions of $a_i$ for $n=0,2,4$ in Figure 5. Thick curves denote the
results for silicates and thin ones are for graphite grains. As expected,
the sputtered fraction decreases with increasing $a_i$, and the fraction
for silicates is in general larger than that for graphites.

The total dust mass fraction that is destroyed can then be evaluated from these
results by convolving $f_d (a_i)$ with an initial grain size distribution.
As mentioned earlier, Nozawa \etal (2003) found that the size distribution
of synthesized grains is bounded by two power laws, with indices
 $-3.5$ for larger
grains and $-2.5$ for smaller grains. We therefore
estimate the total dust mass fraction destroyed, assuming a size
distribution $n(a) da \propto a^{-p} da$, with $p=2.5$ and $3.5$.
The destroyed fraction of dust mass is given by,
\be
f={\int n(a) a^3 f_d(a) da \over \int n(a) a^3 da} \\,
\ee
where the grain volume scales as $a^3$. The results for graphites
and silicates are presented in Table 1, for $n=0,2,4$ 
($E=10^{51}$ erg, $M_{ej}=10^{34}$ g), leaving the cases
of $n > 5$ for which the reverse shocks do not reach the interiors of the
ejecta within the validity range of self-similar solutions.
We assume the lower and upper limit for the integral to be $10^{-7}$ cm
and $3 \times 
10^{-5}$ cm respectively, as explained earlier in \S 2. We also
calculate the destroyed fractions for a truncated grain size distribution,
with an upper limit of $10^{-6}$ cm, and the results are shown inside brackets 
in Table 1 for comparison. 

 We have further studied the variation of the destroyed dust mass fraction
with explosion energy and ejected mass. Tables 2 and 3 show the fraction
of dust mass supttered for two sets of parameters: Table 2 for $M_{ej}=10^{34}$ 
g and $E=10^{52}$ erg, and Table 3 for $M_{ej}=2 \times 10^{34}$ g and
$E=10^{51}$ erg. Figure 6 plots the varition of 
(logarithm) of the destroyed dust mass fraction, $f$, for silicates, for
different values of explosion energy and ejected mass, and for different 
mass profiles ($n=0,2,4$). In the upper panel, we plot the variation of
$f$ with $M_{ej}$ (in logarithm), keeping the explosion energy constant
at $E=10^{51}$ erg, and in the lower palnel, we plot (logarithms of) $f$
with explosion energy $E$, keeping the ejected mass fixed at $M_{ej}10^{34}$
g. The destroyed fraction generally increases with ejected mass,
but its variation with explosion energy is not monotonic. The reason is that
the process of dust destruction depends mainly on two factors: the speed of 
the reverse shock (which depends on the characteristic speed $v_{ch}$) and the
characteristic time, $t_{ch}$. Now, $v_{ch} \propto M_{ej}^{-1/2} E^{1/2}$,
whereas $t_{ch} \propto M_{ej}^{5/6} E^{-1/2}$. It would appear that a
change in ejected mass manifests itself in increasing $t_{ch}$, and thereby
increasing the destroyed fraction, whereas, a change in $E$ affects $t_{ch}$
and $v_{ch}$ in a complicated manner, which results in the variation shown in 
the bottom panel of Figure 6.

\section{Discussion}
We therefore find that a mass fraction
$ \leq 20$ \% of silicates and 
graphite grains
created in cooling
ejecta can be sputtered away by the reverse shock if the
density distribution of the ejecta has a shalow  profile ($n < 5$).
Nozawa et al (2007) recently studied sputtering of dust in reverse shock
and found a destroyed
dust mass fraction of $0.2\hbox{--}1$; their result for 
an ambient hydrogen
number density
of $1$ cm$^{-3}$ and a progenitor mass of $13$ M$_{\odot}$ is $\sim 0.7$
(their Table 1). In fact, their result for the `unmixed grains' model
(which has to do with spatial mixing of the core heavy-element
distribution)
in which the maximum size is $0.01 \, \mu$m
 is $\sim 0.5$.
It should be noted, however, that a considerable amount
of sputtering in their calculation is caused by the hot plasma between
the forward and the reverse shocks (see \S 3.3 in Nozawa et al (2007)),
which we have not considered. In
our case, for analytical simplicity, we have assumed the post-reverse shocked
gas to be cooling adiabatically.
 Our results are, therefore, expected
to give lower bounds on the actual dust mass fraction destroyed
in SN remnants.

There is another reason why the destroyed fraction of dust mass
calculated here is lower than that of Nozawa et al (2007) and Bianchi
\& Schneider (2007). Strictly speaking,
the destruction rate of Tielens et al (1994) used
here is valid for a gas with solar abundance. The gas in the dust
formation region in the supernova ejecta is expected to be metal
rich, and the destruction of dust grains in this gas can be more
efficient in this case, since sputtering yields by metal ions are
much higher in general than by hydrogen and helium ions.

It should be noted that our results for the destroyed fraction is
independent of the ambient density, since the  gas density enters
into our calculation only to change $t_{ch}$ (notice in equations
\ref{eq:char} that $v_{ch}$ is independent of ambient density, and so
the relative speed of reverse shock, and in turn the post-shock temperature,
are also independent of ambient density). In reality, however, a lot
of grain sputtering would take place between the forward and reverse shock,
where the gas density will crucially depend on the ambient density. This
aspect of grain sputtering is neglected in our analytical calculation,
but is captured in the simulation results of Bianchi \& Schneider (2007)
and Nozawa \etal (2007).

Another difference between our results and those of Bianchi \& Schneider 
(2007) arises from the assumption of initial grain size distribution. 
Bianchi \& Schneider (2007) used grains of small size in their 
calculation, with silicate grains smaller than $\le 0.01 \, \mu$m and 
graphites smaller than $\le 0.05 \, \mu$m. Also, their size distribution 
is more confined to a narrow range than the power law distribution assumed
here. Both these factors would enhance the destroyed fraction of dust
mass in reverse shocks. For example, if the sizes of $SiO_2$ grains have
a delta-function distribution around $a_i \sim 25 \, \AA$ (see Figure 1
of Bianchi \& Schneider 2007), then the curves in Figure 2 of
the present paper (say, for $w=0.2$) yield a mass destruction fraction of
$\sim 22$ \% (with $a_f/a_i \sim 0.92$). For $M_gSiO_3$ grains of size
$\sim 45 \, \AA$, we similarly estimate a destroyed mass fraction of
$\sim 14$ \%. From Figure 2, we estimate that one requires $a_I\sim 10 \, \AA$
in order to get  $\sim 50$\% destruction of silicate grains.

A note on the observed density distribution of supernovae ejecta is in
order here.
Chevalier \& Fransson (1994) studied a model of the ejecta
from a Type II supernova  with a shallow ($n<3$) core surrounded by a
steep ($n>5$) envelope. In the case of Type Ia supernovae, a steep density
distribution in the ejecta mass has also been discussed in the literature
(Colgate \& McKee (1969); Dwarakadas \& Chevalier (1997)).
There has been success, however, in modelling
data of Type Ia supernova with uniformly distributed ($n\sim0$) ejecta
(Hamilton \& Sarazin (1984); Hamilton, Sarazin \& Szymkowiak (1986)).

It is worth considering what observations might improve our
understanding of the net dust production by SNe.  Far-infrared and
sub-millimeter observations of reverse-shocked dust in young SNe
by {\it Spitzer}, {\it Herschel}, and other facilities will help,
but our theoretical studies suggest that unshocked (cold) dust
may harbor a considerable mass of undetected dust. To constrain
this dust and distinguish it from surrounding interstellar clouds
will probably require sub-mm observations with  small beam sizes.
This brings in complications of the circumstellar environment of
young SNe.  Furthermore, the long times (thousands of years) required
for reverse shocks to reach the core ejecta suggest that late-time
observations of supernova remnants would be useful.

\section{Summary}
We have studied the effect of reverse shocks analytically,
in the regime of self-similar
evolution, on the sputtering of
dust grains that are believed to form in cooling ejecta. For representative
cases we found that fractions of dust mass that is destroyed are of order
of order $1\hbox{--}20$ \% of silicates and 
graphites,
for SN of explosion energy $ \sim 10^{51}$ erg, ejecta mass
$\sim 10^{34}$ g, and an ambient  density $ \sim 10^{-24}$ g  cm$^{-3}$.
Our analytical formalism provides only a lower bound on the dust mass
fraction that is destroyed by the reverse shock,
since it ignores further sputtering of grains
in hot plasma between the forward and reverse shocks. Our results are,
therefore,
 consistent
with the recent estimates from numerical simulations which
include these additional effects.
Furthermore, Our study provides a formalism to generalize these results
to cases with different values of SN parameters.

\acknowledgements We thank Drs. R. Chevalier, E. Dwek, T. Nozawa, A. Ray
and Y. Shchekinov for helpful discussions.
This work was supported at the University of Colorado by NASA Theory
grant NNX07-AG77G and NSF theory grant AST07-07474.

\clearpage

\begin{table}
\centerline{Table 1 : Fraction of dust mass sputtered for sizes ranging from
$10^{-4} \, \mu$m to $a_{max}=0.3 \, \mu$m (or $a_{max}=0.01 \, \mu$m)
}
\vskip 0.2in
\begin{center}
\begin{tabular}{|c||c||c|c|}
\hline
grain ($dn/da$) & $n=0 $ & $n=2$ & $n=4$  \\
\hline
\hline
Graphite ($\propto a^{-3.5}$) & $0.004 \, (0.02) $ & $0.01 \,(0.05) $ & $0.007 \, 
(0.04)$  \\
\hline
($\propto a^{-2.5}$) & $6 \times 10^{-4} \,(0.01)$ & $0.001 \,(0.03) $ & $0.001 \,(0.03) $ \\
\hline
Silicate ($\propto a^{-3.5}$)& $0.01 \,(0.07) $ & $0.03 \, (0.13) $ & $0.02 \,(0.10)$  \\
\hline
($\propto a^{-2.5}$) & $0.002 \, (0.05) $ & $0.005 \, (0.10) $ & $0.004 \, (0.07)$  \\
\hline\hline
\end{tabular}
\end{center}
\end{table}

\begin{table}
\centerline{Table 2 : Fraction of dust mass sputtered  for $M_{ej}= 10^{34}$
g, $E_{ej}=10^{52}$ erg.
}
\vskip 0.2in
\begin{center}
\begin{tabular}{|c||c||c|c|}
\hline
grain ($dn/da$) & $n=0 $ & $n=2$ & $n=4$  \\
\hline
\hline
Graphite ($\propto a^{-3.5}$) & $0.001 \, (0.008) $ & $0.003 \,(0.01) $ & $0.005 \, 
(0.03)$  \\
\hline
($\propto a^{-2.5}$) & $2 \times 10^{-4} \,(0.005)$ & $4 \times 10^{-4} \,(0.01) $ & $7 \times 10^{-4} \,(0.02) $ \\
\hline
Silicate ($\propto a^{-3.5}$)& $0.006 \,(0.03) $ & $0.01 \, (0.06) $ & $0.05 \,(0.16)$  \\
\hline
($\propto a^{-2.5}$) & $0.001 \, (0.02) $ & $0.002 \, (0.04) $ & $0.018 \, (0.14)$  \\
\hline\hline
\end{tabular}
\end{center}
\end{table}

\begin{table}
\centerline{Table 3 : Fraction of dust mass sputtered  for $M_{ej}= 
2 \times 10^{34}$ g,
 $E_{ej}=10^{51}$ erg.
}
\vskip 0.2in
\begin{center}
\begin{tabular}{|c||c||c|c|}
\hline
grain ($dn/da$) & $n=0 $ & $n=2$ & $n=4$  \\
\hline
\hline
Graphite ($\propto a^{-3.5}$) & $0.006 \, (0.035) $ & $0.015 \,(0.08) $ & $0.01 \, 
(0.06)$  \\
\hline
($\propto a^{-2.5}$) & $10^{-3} \,(0.02)$ & $0.003 \,(0.06) $ & $0.002 \,(0.04) $ \\
\hline
Silicate ($\propto a^{-3.5}$)& $0.02 \,(0.1) $ & $0.04 \, (0.18) $ & $0.03 \,(0.13)$  \\
\hline
($\propto a^{-2.5}$) & $0.003 \, (0.07) $ & $0.008 \, (0.14) $ & $0.005 \, (0.10)$  \\
\hline\hline
\end{tabular}
\end{center}
\end{table}

\clearpage

\begin{figure}
\epsscale{1.0}
\plotone{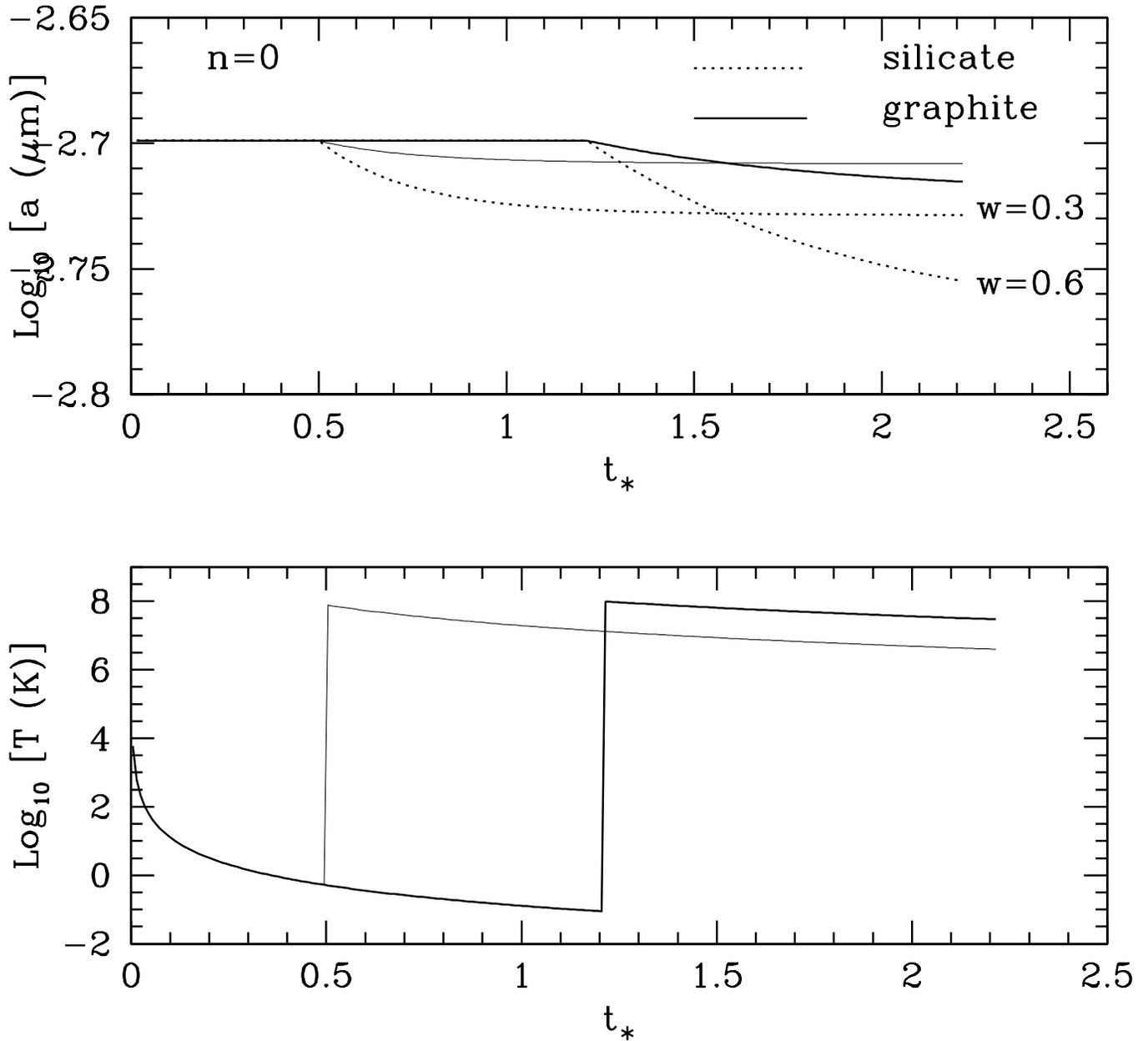}
\caption{
The dust grain size as a function of time (where $t_\ast=t/t_{ch}$) is plotted
for $n=0$, $E=10^{51}$ erg, $M_{ej}=10^{34}$ g
in the upper panel, for two shells, with $w=0.6$
and $w=0.3$. The shell $w=0.3$ is inside of $w=0.6$ shell, and is hit by the
reverse shock at a later time. The initial grain size (solid line for graphites,
and dotted, for silicates) is $1.3 \times 10^{-3}$ $\mu m$.
The bottom panel plots gas temperature as a function of $t^\ast$ showing
the rise at the instant of the reverse shock meeting the two particular shells.
($t_{ch}\sim 1.7 \times 10^3 n_0^{-1/3}$ yr.)
}
\end{figure}

\clearpage

\begin{figure}
\epsscale{1.0}
\plotone{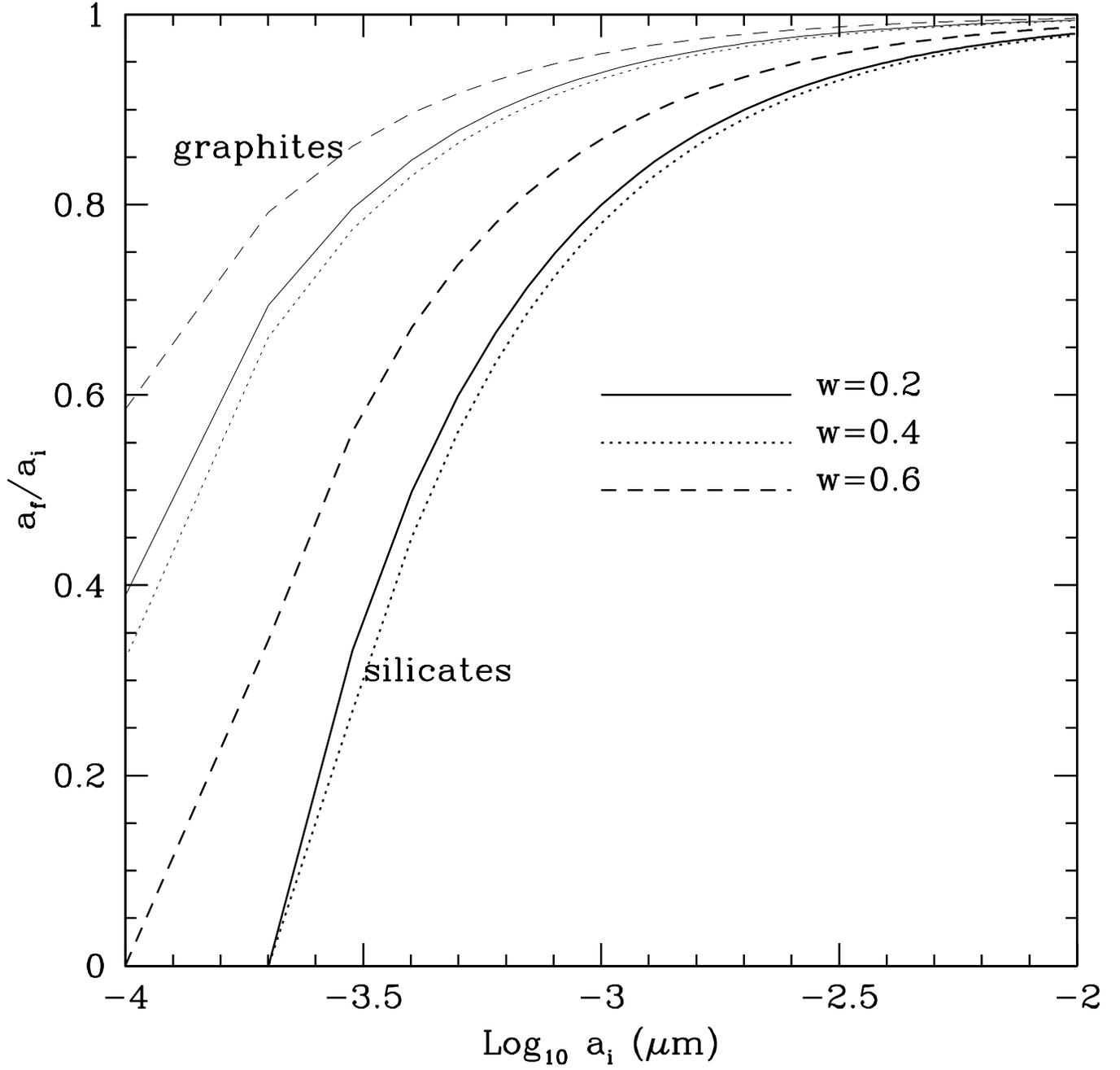}
\caption{
Ratio of final to initial grain radius is plotted against initial grain
radius (in $\mu$m) for silicates (thick lines) and graphites (thin lines),
for grains in shells $w=0.2$ (solid),
 $w=0.4$ (dotted), $w=0.6$ (dashed) for the case
of $n=0$.
}
\end{figure}

\clearpage

\begin{figure}
\epsscale{1.0}
\plotone{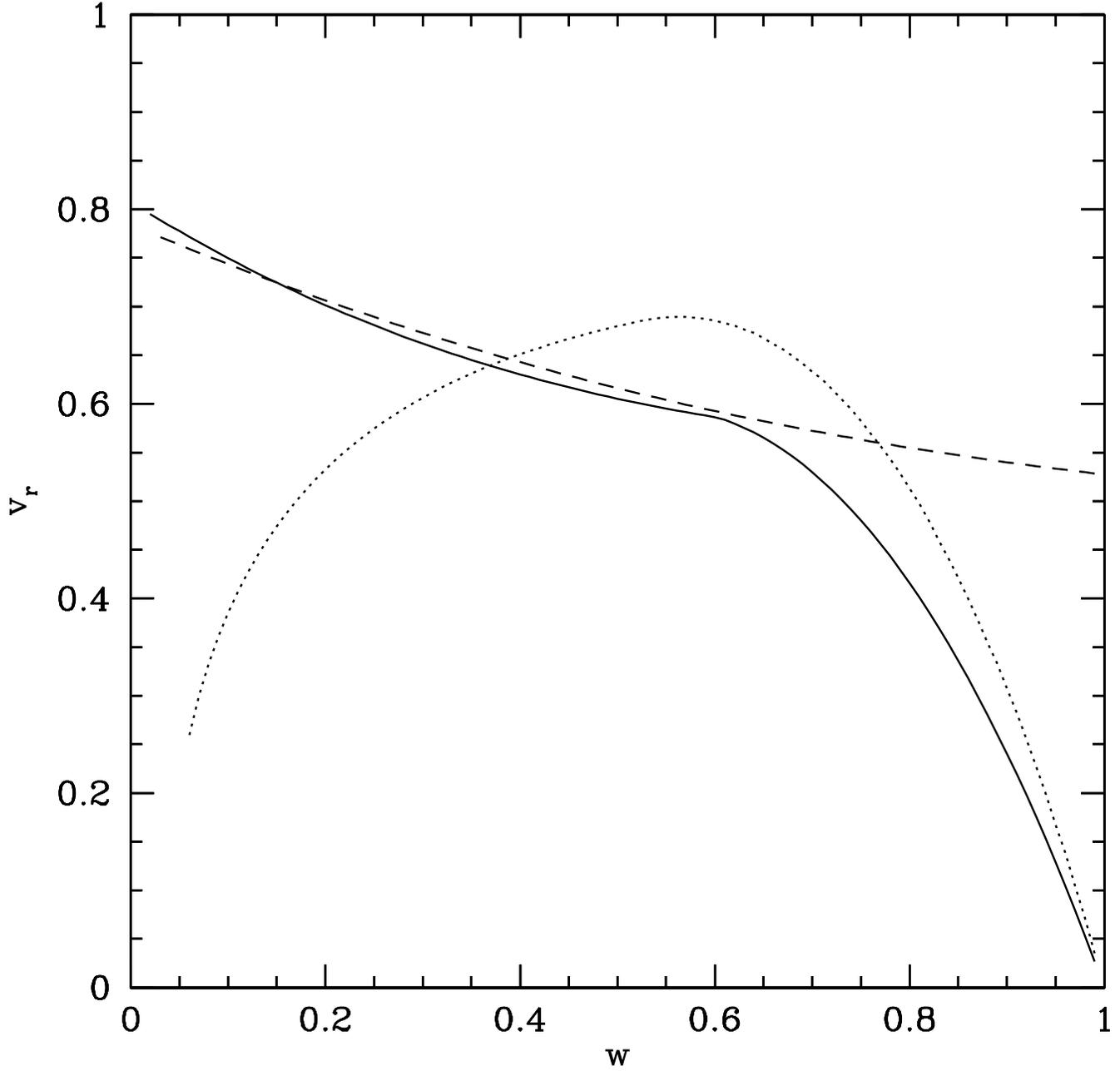}
\caption{
The dimensionless velocity of reverse shock in the frame of unshocked ejecta,
$\tilde{v_r} ^\ast$
(in the units of the characteristic velocity $v_{ch}$)
is plotted against $w$ for $n=0,2,6$ (solid, dotted,
dashed lines). For reference, the value of $v_{ch} \sim 3162$ km
s$^{-1}$, for $E=10^{51}$ erg and $M_{ej}=10^{34}$ g.
}
\end{figure}

\clearpage

\begin{figure}
\epsscale{1.0}
\plotone{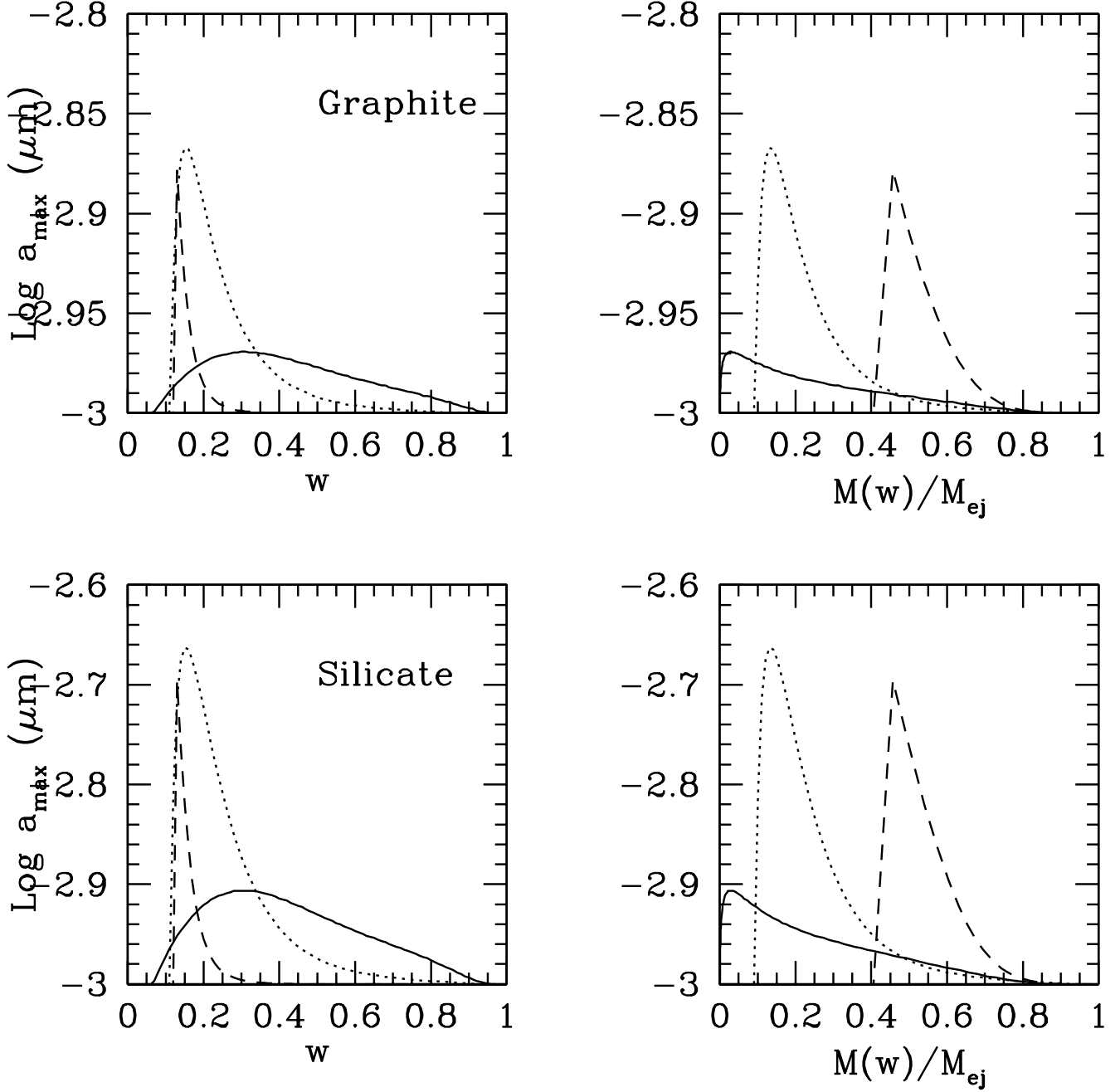}
\caption{
The values of maximum size, $\log a_{max}$ (in $\mu$m), 
below which  grains are  sputtered
below $10^{-7}$ cm
 are shown against the parameter $w$ in the left panels, and against
 the fraction of ejecta mass inside
the corresponding shell, $M(<w)/M_{ej}$, in the right panels.
($w=R(t)/(v_{ej}t)$, or equivalently, it
 is the ratio of the velocity of the ejecta in the shell to the
maximum ejecta velocity.) The
case of graphites are shown in upper panels and the case of silicates,
in the lower panels. Dust
is introduced at 10 yr, where the tracks begin. Solid, dotted, and dashed
lines refer to the cases $n=0,2,4$ respectively.
The energy of explosion is assumed to be $E_{ej}=10^{51}$ erg and ejecta
mass is assumed to be $10^{34}$ g.
}
\end{figure}

\clearpage

\begin{figure}
\epsscale{1.0}
\plotone{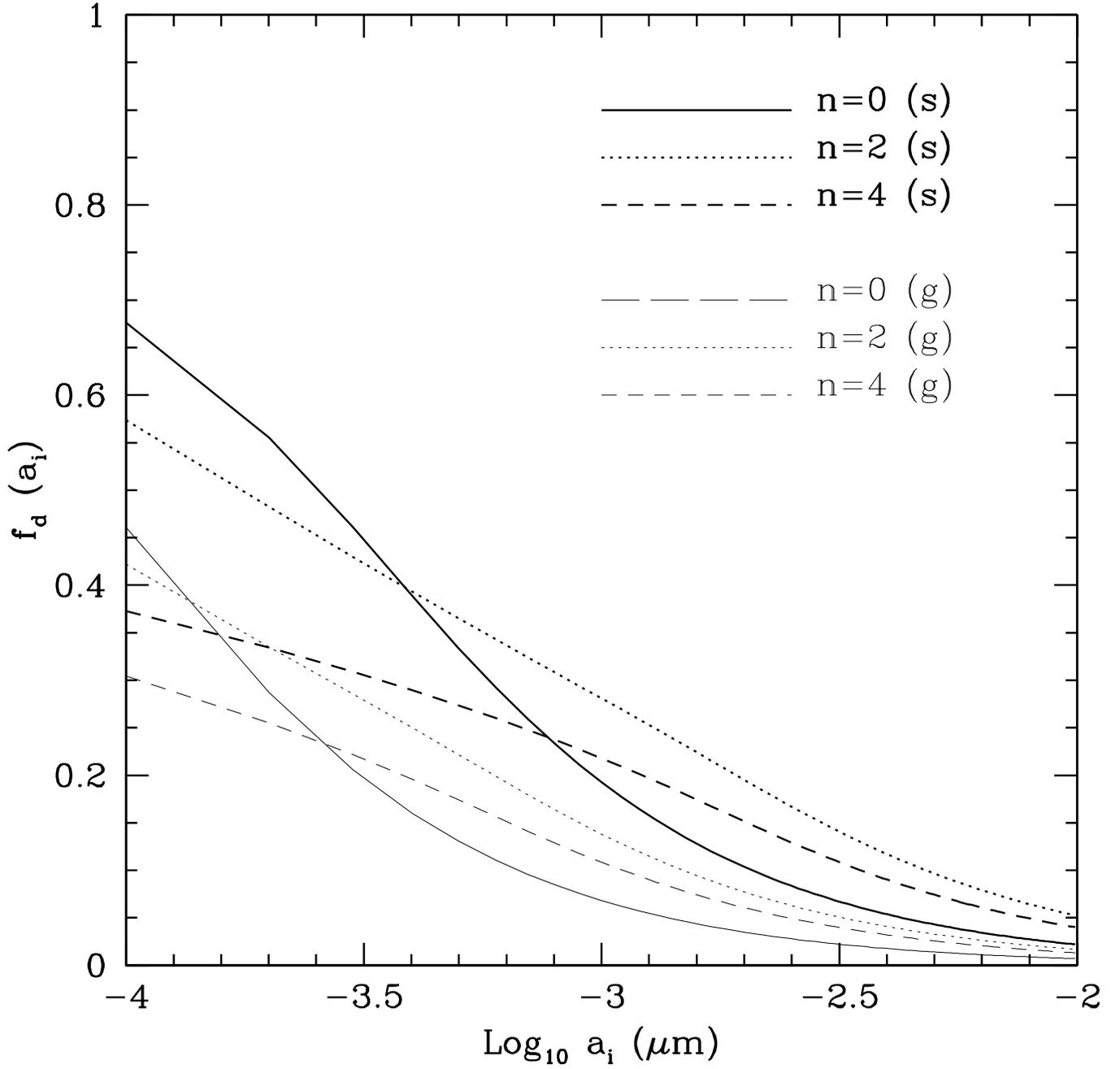}
\caption{
Fraction of dust mass sputtered as a function of initial grain size,
weighted by the mass of the shell in which they are formed,
for $n=0$ (solid), $2$ (dotted) and $4$ (dashed), for silicates
(thick curves) and graphites (thin curves).
}
\end{figure}

\clearpage

\begin{figure}
\epsscale{1.0}
\plotone{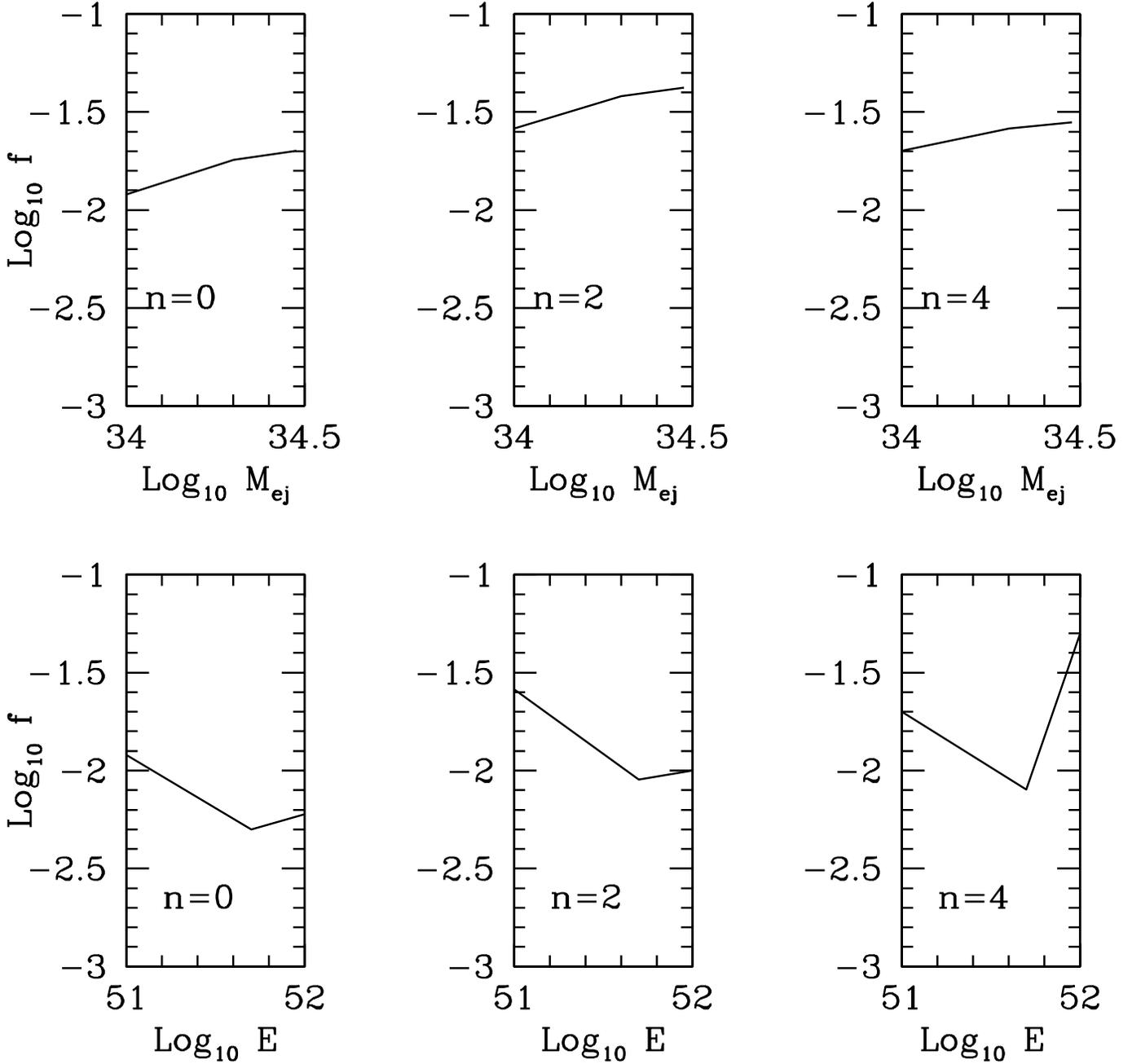}
\caption{
Fraction of dust mass sputtered is plotted in the upper
panel as a function of mass of ejecta, keeping the explosion energy
constant at $10^{51}$ erg, for different values of $n$. In the lower
panel, the fraction of destroyed dust mass is plotted (for different $n$)
as a function of explosion energy $E$, keeping the ejecta mass
constant at $10^{34}$ g. 
}
\end{figure}

\clearpage

\end{document}